%% file: example_paper.tex
\documentclass[9pt,sigconf]{acmart}

\renewcommand\footnotetextcopyrightpermission[1]{} 

\usepackage{soul}
\usepackage{xcolor}
\usepackage{algorithm}
\usepackage{algorithmic}
\usepackage{dsfont}
\usepackage{graphicx}
\usepackage{amsmath}
\usepackage{textcomp}
\usepackage{subcaption}



\begin{document}

\title{Efficient Task Transfer for HLS DSE}

\author{Zijian Ding, Atefeh Sohrabizadeh, Weikai Li, Zongyue Qin, Yizhou Sun, Jason Cong}
\email{{bradyd,atefehsz,weikaili,qinzongyue,yzsun,cong}@cs.ucla.edu}
\affiliation{
  \institution{University of California, Los Angeles}
  \city{Los Angeles}
  \state{CA}
  \country{USA}
  \postcode{90095}
}

\begin{abstract}
There have been several recent works proposed to utilize model-based optimization methods to improve the productivity of using high-level synthesis (HLS) to design domain-specific architectures. They would replace the time-consuming performance estimation or simulation of design with a proxy model, and automatically insert pragmas to guide hardware optimizations. In this work, we address the challenges associated with high-level synthesis (HLS) design space exploration (DSE) through the evolving landscape of HLS tools. As these tools develop, the quality of results (QoR) from synthesis can vary significantly, complicating the maintenance of optimal design strategies across different toolchains. We introduce Active-CEM, a task transfer learning scheme that leverages a model-based explorer designed to adapt efficiently to changes in toolchains. This approach optimizes sample efficiency by identifying high-quality design configurations under a new toolchain without requiring extensive re-evaluation. We further refine our methodology by incorporating toolchain-invariant modeling. This allows us to predict QoR changes more accurately despite shifts in the black-box implementation of the toolchains. Experiment results on the HLSyn benchmark transitioning to new toolchain show an average performance improvement of 1.58$\times$ compared to AutoDSE and a 1.2$\times$ improvement over HARP, while also increasing the sample efficiency by 5.26$\times$, and reducing the runtime by 2.7$\times$.
\end{abstract}

\maketitle
\pagestyle{plain}

\input{1_intro.tex}
\input{2_preliminaries.tex}
\input{4_transfer.tex}
\input{5_search.tex}
\input{6_evaluation.tex}

\section{Conclusion}
In this work, we discuss the sample efficiency challenge in the task transfer learning problem for HLS DSE. We mitigate the challenge with a novel approach on both sampling and modeling. On sampling, we propose Active-CEM, a model-based explorer that efficiently samples high-quality designs in the whole design space. On modeling, we propose to model both the invariance and difference between toolchains and jointly train a single model with data collected from multiple toolchains. Moving forward, we plan to evaluate our sampling method in the domain-transfer setting and consider task-transfer learning for different devices. We also plan to combine domain knowledge in the sampling process for better efficiency.

\begin{acks}
This work was partially supported by NSF grants 2211557, 1937599,  2119643, and 2303037, SRC JUMP 2.0 PRISM Center, NASA, Okawa Foundation, Amazon Research, Cisco, Picsart, Snapchat, and the CDSC industrial partners (https://cdsc.ucla.edu/partners/). The authors would also like to thank AMD/Xilinx for HACC equipment donation, and Marci Baun for editing the paper. J. Cong has a financial interest in AMD.
\end{acks}

\clearpage
\newpage
\bibliographystyle{ACM-Reference-Format}
\bibliography{example_paper}

\newpage
\appendix
\input{7_appendix}

\end{document}

%% file: 1_intro.tex
\section{Introduction}
\label{sec:intro}

General-purpose computers are widely employed across diverse domains owing to their ease of programming. However, they face significant overheads due to extended instruction pipelines necessary for supporting generalization. 
 
This has given rise to domain-specific accelerators (DSAs)~\cite{dally2020domain, chi2022democratizing, cong2019customizable}. However, DSAs pose challenges in programming, restricting their user base to hardware experts~\cite{chi2022democratizing}. High-level synthesis (HLS)~\cite{cong2011high} was introduced to simplify the design process by raising the abstraction level from the register-transfer level (RTL) to C/C++.  Designers use compiler directives, in the form of pragmas, to describe microarchitecture. While HLS reduces design turn-around times, not every HLS design yields a high-performance microarchitecture~\cite{sohrabizadeh2022autodse}. As a result, a new line of research aims to explore the solution space defined by pragmas more efficiently. Because evaluating each design candidate with HLS tools is time-consuming, recent studies propose supervised learning models as proxies of HLS tools, which can predict the quality of results (QoR) to expedite exploration~\cite{ustun2020accurate, wu2022ironman, bai2022improving, sohrabizadeh2022automated, wu2022high}. Training such models involves collecting extensive datasets of diverse designs synthesized by HLS tools.

\begin{figure}
    \centering
    \subfloat[V20 vs. V21]{
        \includegraphics[height=0.11\textheight]{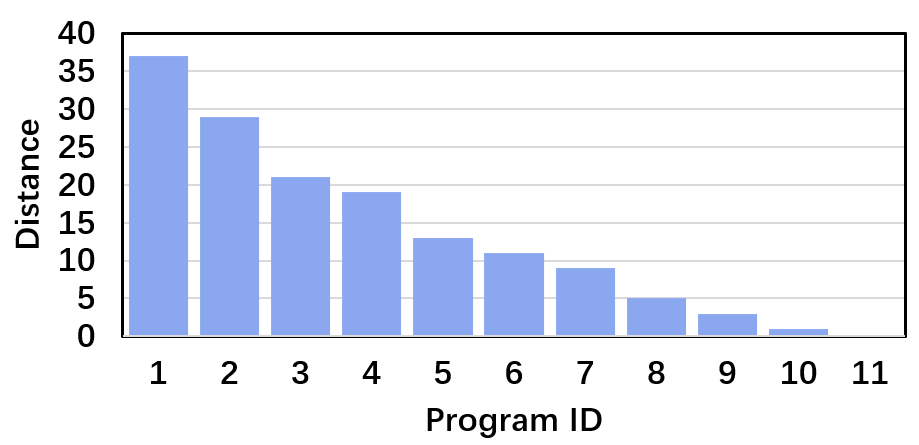}
        \label{fig:editing_v21}
    }
    \subfloat[V21 vs. V23]{
        \includegraphics[height=0.11\textheight]{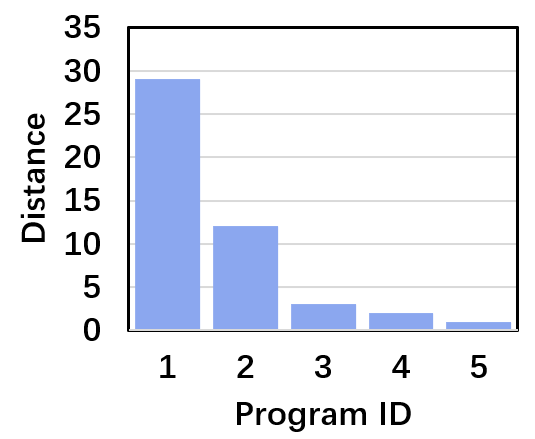}
        \label{fig:editing_v23}
    }
    \caption{The distance of the best design found by AutoDSE~\cite{sohrabizadeh2022autodse} between different toolchains: The horizontal axis represents different programs, and the vertical axis denotes the distance between designs. V20: Vitis HLS 2020.2, V21: Vitis HLS 2021.1, V23: Vitis HLS 2023.2. The distance is calculated by summing up the discrete code difference between two designs over each pragma.}
    \label{fig:editing}
\end{figure}

A significant issue arises from the continuous evolution of HLS tools, which can significantly impact QoR values~\cite{sohrabizadeh2023robust}. To gain insight into this issue, we utilized AutoDSE~\cite{sohrabizadeh2022autodse} to gather data on different Vitis HLS toolchains: 2020.2 (V20)~\cite{vitis20}, 2021.1 (V21)~\cite{vitis21} and 2023.2 (V23)~\cite{vitis23}. 

\begin{figure}
    \centering
     \subfloat[LUT-V20 vs. V21]{
        \includegraphics[width=0.15\textwidth]{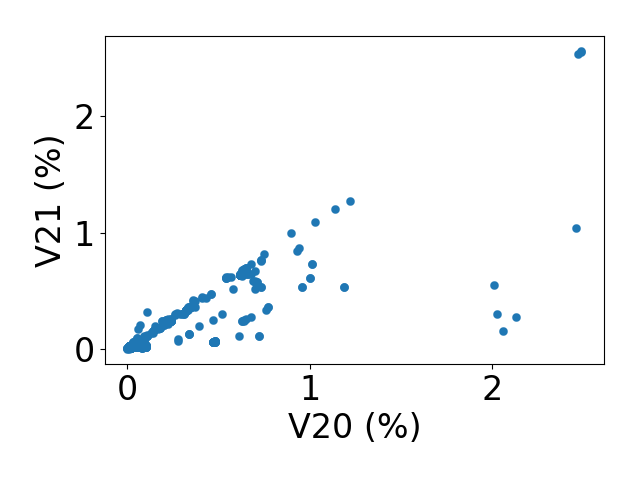}
    }
    \subfloat[LUT-V21 vs. V23]{
        \includegraphics[width=0.15\textwidth]{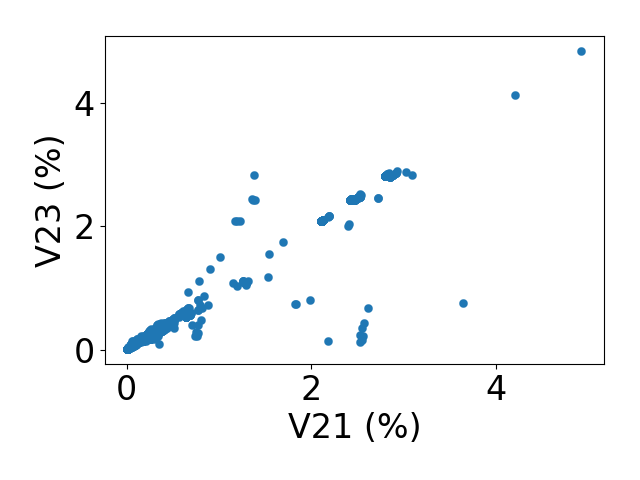}
    }
    \subfloat[Latency-V20 vs. V21]{
        \includegraphics[width=0.15\textwidth]{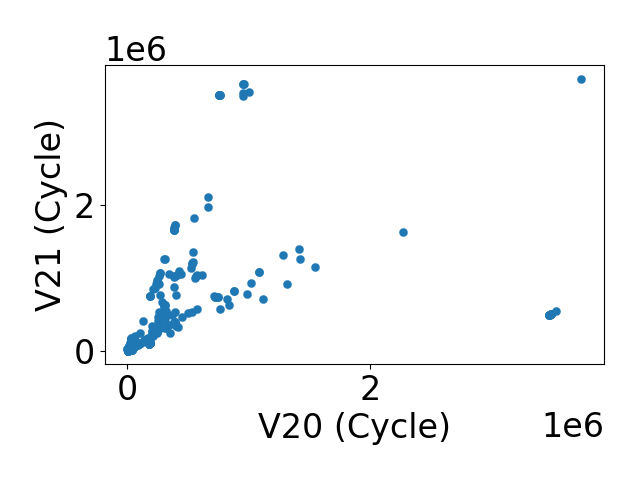}
    }
    \caption{The label shift when changing the toolchain: the performance/resource labels are plotted on the X/Y axis for the old/new toolchain, a larger distance from the $Y=X$ line indicate larger label shift.}
    \label{fig:label_shift}
\end{figure}

Figure \ref{fig:label_shift} illustrates how the performance and resource utilization of each design may vary when changing the toolchain. Additionally, as shown in Table \ref{tab:valid_shift}, the validity of a design may also fluctuate. During HLS design synthesis, encountering errors is common. For instance, the HLS tool may refuse to synthesize a design with high parallelization factors~\cite{sohrabizadeh2022automated}, or it may timeout during synthesis. It is worth noting  that we only consider designs within the intersection of those explored by AutoDSE, and more significant changes may occur when evaluating designs across the entire design space. One question that arises is whether changes in performance/resource labels and validity lead to alterations in the optimal design scheme. For instance, a design that previously performed well with one toolchain may become invalid with another. To understand how changes in HLS tools affect the design scheme, we run AutoDSE on these toolchains and compare the best design it finds. AutoDSE optimizes design performance based on feedback from the toolchain and can naturally adapt to new toolchains. Therefore, any change in the best design found by AutoDSE is attributed to the change in the downstream toolchain. Upon applying AutoDSE to the new toolchain, we observed an average performance improvement of $2\times$. Furthermore, we noted that the change in the performance of the best design arises from the modification in the design scheme. Essentially, AutoDSE can discover new and improved designs on the new toolchain. Figure \ref{fig:editing} depicts the editing distance of the best designs found by AutoDSE on the new and old toolchains. The pragma edit distance is calculated by summing up the differences between each pragma. For example, if one pragma $P_1$ can take values from [1, 2, 5], and one design takes value 5 while the other takes value 1, then their discretized values (index in the design space) are 3 and 1 respectively. This results in a contribution of $3-1=2$ to the total editing distance. A larger editing distance indicates a greater shift in the design scheme.

\begin{table}
\caption{The validity changes when altering the toolchain. We compare two pairs of commercial HLS tools: V20 vs. V21, V21 vs. V23. ``T'' and ``F'' stand for the valid or invalid design in the corresponding toolchain.}
\label{tab:valid_shift}
\centering
\subfloat[V20 vs. V21]{
\begin{tabular}{|c|cc|}
\hline
            & V21 T & V21 F \\
\hline
V20 T   & 1462 & 295 \\
\hline
V20 F & 269  & 348 \\
\hline
\end{tabular}
}
\subfloat[V21 vs. V23]{
\begin{tabular}{|c|cc|}
\hline
            & V23 T & V23 F \\
\hline
V21 T   & 3690 & 120 \\
\hline
V21 F & 72   & 1814 \\
\hline
\end{tabular}
}
\end{table}

It is therefore critical to design a task transfer learning scheme that could efficiently adapt to the evolving toolchain. According to the previous observations, we observe two major challenges. (1) The first one is to sample efficiently the design from the design space so that the model transferred to those designs can effectively identify the high-quality designs. (2) Another challenge is to adapt the model to the large label shift. Overall, we want to efficiently uncover good designs on the new toolchain when the label shift is unknown and the design space could be as large as $10^{13}$ as we see in the HLSyn benchmark~\cite{bai2023towards}.

Model-based optimization is a widely used technique for efficiently sampling high-quality designs. Previous work has utilized Bayesian Optimization~\cite{budak2023practical,sun2022correlated}, where sampling is guided by an acquisition function. However, optimizing the acquisition function in a large discrete space, such as HLS designs, remains challenging. Other studies, for example~\cite{lin2018data} have adapted active learning to achieve better sample efficiency when transferring to a new technology node by proposing data selection algorithms that select diverse and informative data from existing datasets. Our approach is not limited to selecting data from an existing unlabeled data pool, and could navigate the whole design space. In ~\cite{wu2022high, wu2022ironman, sohrabizadeh2022automated, sohrabizadeh2023robust, ustun2020accurate}, optimizing HLS designs involves training a proxy model on an offline dataset and then conducting optimization with this proxy model. This offline approach requires time-consuming data collection and suffers from generalizability issues when applied to new domains and tasks.

Transfer learning investigates the generalizability of models to unseen domains and tasks, with representation learning playing a critical role in domain transferability~\cite{ganin2016domain,AdaGCN,DANE,wu2022eerm,zhu2020transfer,GraphMETRO, bai2022improving}. Some studies propose learned design space pruning for domain transfer in Bayesian Optimization~\cite{perrone2019learning, bai2023transfer}. Unlike domain transfer, which assumes a data distribution shift, we focus on task transfer, assuming a tractable change in the black-box function. Previous works~\cite{sohrabizadeh2023robust, lin2018data} have shown better task transfer accuracy by freezing some model parameters.

To enable efficient sampling on an unseen toolchain, we propose a novel model-based explorer designed for discrete optimization spaces. We jointly optimize the model prediction and the importance sampling distribution, allowing the sampling process to leverage knowledge from the previous toolchain and achieve better sample efficiency. Moreover, the sampling distribution can guide model updates, enabling the model to minimize its error on a subset of the design space rather than striving for perfect accuracy across the entire space. We also explore learning task-generalizable representations of HLS designs. Our key observation is that the same input often represents similar micro-architectures, despite potential differences in their implementations across toolchains. Thus, we aim to learn representations of HLS designs that can extract their invariance across various toolchains.

Overall, we make the following contributions:

\begin{itemize}
\item We propose a model-based explorer that efficiently searches the whole design space, with better sample efficiency compared with the SoTA approach.
\item We introduce a novel modeling of HLS designs to capture the invariance between toolchains.
\item Evaluation on 40 programs transferring to V21 shows that we can improve design performance by  1.58$\times$ and 1.2$\times$ on average, compared with AutoDSE and HARP, with a sample efficiency of 5.26$\times$, and a 2.7$\times$ runtime improvement.
\item Evaluation on the latest toolchain reveals that our design outperforms the best of AutoDSE and HARP by 1.27$\times$.
\end{itemize}

%% file: 2_preliminaries.tex
\section{Preliminaries}

\subsection{HLS Design Space and Pragmas} 
\label{sec:bg_hls}
As with AutoDSE and HARP, we utilize the open-source AMD/Xilinx Merlin Compiler~\cite{cong2016source} to streamline DSE for HLS. It supports three types of pragmas: PIPELINE, PARALLEL, and TILE. Taking designs with these pragmas as input, it automates the synthesis of on-chip buffers and memory interfaces. The PIPELINE, PARALLEL, and TILE pragmas handle loop pipelining, double buffering, loop unrolling, and loop tiling for parallelization and latency hiding. The Merlin Compiler supports standard pipelining and parallelization of HLS, but also offers additional automation for pragmas like ``array\_partition" which can be optimally solved given the parallel and tiling factors. This compiler provides a more compact design space and is used in this work. However, our approach can be directly applied to other HLS tools as well.

\subsection{HARP}
The HARP framework~\cite{sohrabizadeh2023robust} is the state-of-the-art model-based approach for predicting the performance and resource utilization of HLS designs. It introduces a hierarchical graph representation to mitigate the over-smoothing issue encountered in graph neural networks. By employing a hierarchical graph, the shortest path of the graph can be substantially reduced. This enables the capture of more global information with a shallow GNN. Furthermore, HARP proposes to model the program and pragma transformations separately, treating pragmas as transformations modeled by Multi-Layer Perceptrons (MLPs). This model design aligns with the nature of HLS designs and results in improved accuracy and DSE performance. It trains a classification model to predict the validity and a regression model to predict the QoR. We follow this convention and denote the classification model as $\hat{R}_{\theta_c}$ and the regression model as $\hat{R}_{\theta_r}$.

%% file: 4_transfer.tex
\section{Methodology}

\begin{figure}
    \centering
\includegraphics[width=0.75\columnwidth]{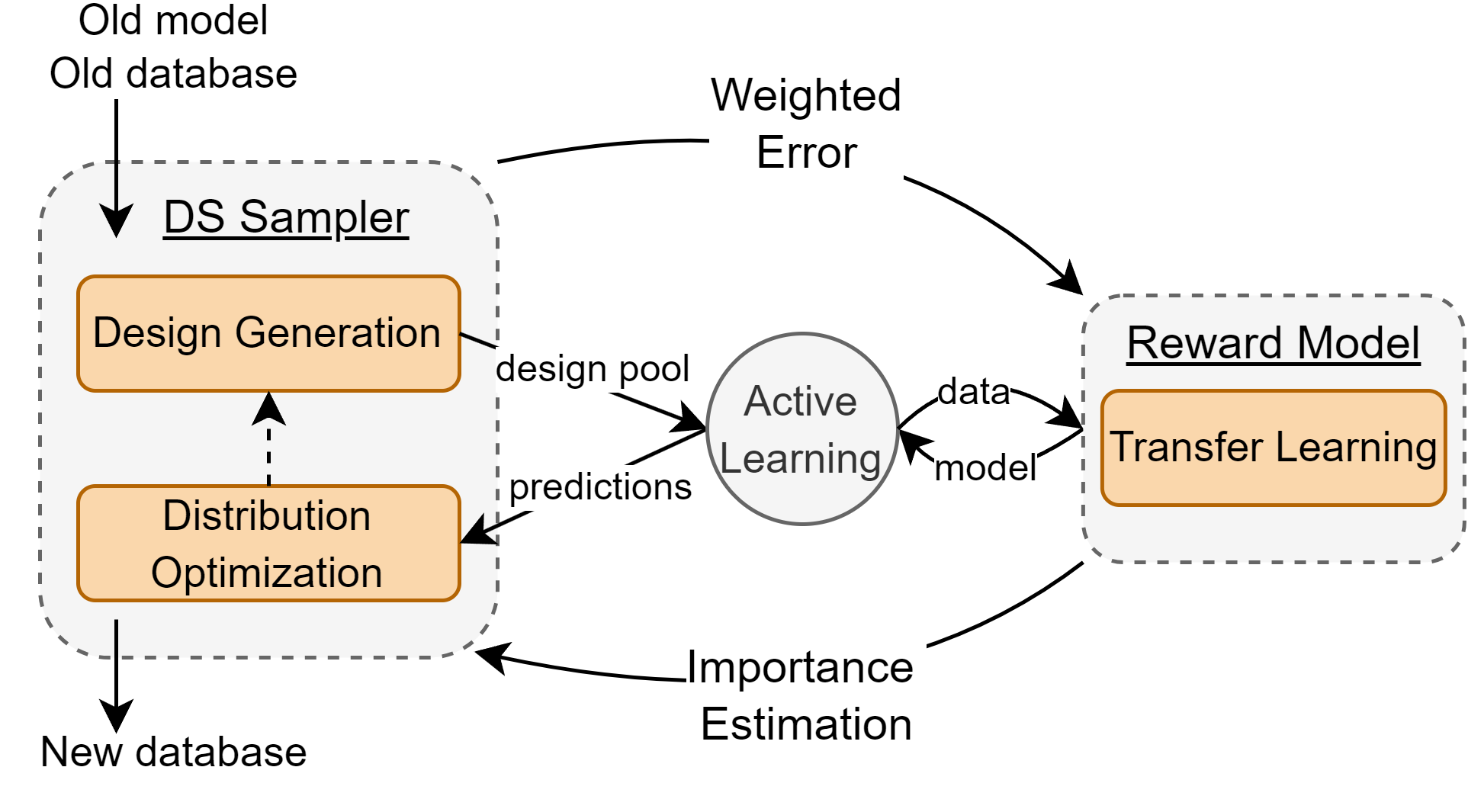}
    \caption{Overview: The DS Sampler and the Reward Model interact with each other through active learning}
    \label{fig:overview}
\end{figure}

Our primary objective is for the learned model to adapt to a new toolchain while identifying Pareto-optimal points and maximizing sample efficiency. Our approach focuses on employing importance sampling to select high-quality designs and integrating model prediction within the importance sampling algorithm to improve sample efficiency. Figure \ref{fig:overview} illustrates an overview of our proposed method, which consists of two main components: the DS Sampler and Reward Model Transfer Learning. Using the Cross-Entropy Method (CEM), the DS Sampler performs importance sampling based on the design performance. It iteratively optimizes a probability distribution across the entire design space, concentrating higher density on designs with high rewards. This approach ensures that high-reward designs are sampled more frequently, thereby increasing the likelihood of identifying optimal designs. To further boost sample efficiency, we introduce a reward model that predicts the validity and QoR of HLS designs. This model is adapted from previous toolchains, leveraging data collected from earlier implementations to enhance prediction accuracy. To address the potential label shift due to toolchain changes and distribution shift from the importance sampling, we integrate an active learning subroutine within the CEM algorithm. In summary, we facilitate an efficient transfer to a new toolchain through a model-based method, which involves two iterative steps: fitting the model to the current sampling distribution and updating the sampling distribution based on the model's recommendations.

In Section \ref{sec:transfer}, we will describe version-invariant modeling that efficiently adapts the model to the label shift. In Section \ref{sec:search}, we will explain the model-based importance sampling algorithm based on CEM.

\subsection{Toolchain-invariant modeling}
\label{sec:transfer}

A critical problem of training a prediction model for HLS designs is how to train a robust model when the labeled data is scarce compared to the size of the design space. One way to mitigate the insufficiency of labeled data is to transfer knowledge from all existing data, possibly collected from different toolchains. Our intuition is to model a robust representation of HLS designs that contains the invariant information between different toolchains. In practice, such invariance is abundant. For example, the same input program with the same pragmas represents similar microarchitectures, and similar microarchitectures will not result in completely different implementations. Following this intuition, we consider learning invariant embedding for task transfer on the HLS design by training a single model with data from different toolchains.

\begin{figure}
    \centering
    \includegraphics[height=0.15\textheight]{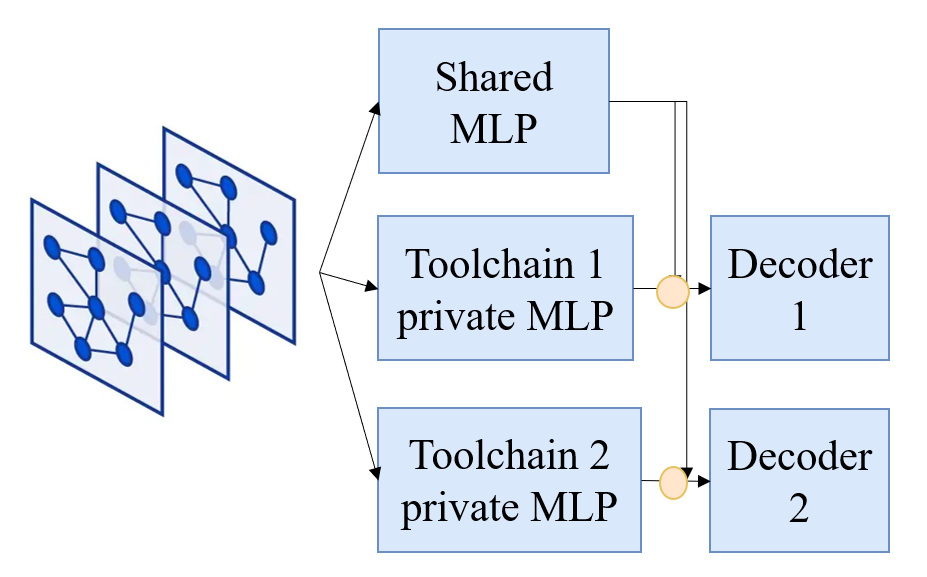}
    \caption{Model architecture with invariant embedding}
    \label{fig:inv_hybrid}
\end{figure}

One problem to tackle is to train a model that is aware of the version of the data. A simple solution is to include the version tag in the input data. However, when switching to a new version, we may need a completely new tag, and that will cause a huge shift in input data. Another approach is to have a shared encoder, but different decoder for each task. Each decoder is trained on data from its corresponding toolchain, allowing the unique parameters of the decoders to capture the specific characteristics of each toolchain. This approach follows the intuition that the same input HLS design represents a similar micro-architecture for different toolchains, and could be represented as the same embedding in the latent space. However, we argue that it is not enough to just model the shared information between toolchains. A shared embedding may not contain enough information to address the variance between toolchains. In HLS design, the information that remains completely invariant is the program. But for pragma transformations, it remains unknown exactly what implementation stays the same and what is changed. Following this intuition, we propose the following modeling scheme:

\textbf{Separate embedding for shared and private information.} Fig. \ref{fig:inv_hybrid} shows the model architecture. For each input design, two separate embeddings of the design will be learned. One of the embeddings is modeled by a set of parameters shared among all toolchains, and the other embedding is modeled with a set of parameters that is private to each toolchain. Then, we concatenate the two embeddings and forward it to different decoders for each toolchain.

Another benefit of the proposed model architecture is that it can be trained jointly with data from multiple versions. We optimize the mean-squared error (MSE) averaged over all versions of data. Intuitively, training with more data will help the model get a stronger representation of the invariant part. Moreover, once diverse data from different toolchains have been gathered, the invariant part can be pretrained, and we are able to fix the invariant part and fine-tune only the private part when switching to a new toolchain. This will result in fine-tuning fewer parameters when transferring to a new toolchain and better efficiency.

%% file: 5_search.tex
\section{DS Sampler}
\label{sec:search}

\subsection{Optimization with Cross-Entropy Method}
\label{sec:cem_free}
We will now describe the original model-free version of the Cross-Entropy Method (CEM)~\cite{de2005tutorial} and how it can be applied to optimize HLS design. It was originally designed for rare-event simulation but can be applied for global optimization, and it has the ability to converge to global optimum theoretically~\cite{rubinstein2004cross, margolin2005convergence}. While the cross-entropy method is treated as a model-based technique in some work~\cite{costa2007convergence} because it contains a model for the distribution of rare events (good designs), we treat the original version of the algorithm as model-free since it does not contain a learned model of the reward function. We will show that the cross-entropy method is a strong candidate for efficient global optimization.

\begin{algorithm}[tb]
   \caption{Cross-entropy method for optimization (model-free)}
   \label{alg:cem_free}
\begin{algorithmic}[1]
   \STATE {\bfseries Input:} distribution hypothesis set $\Theta$, number of iterations $n$, sample per iteration $N$, true function $R$, quantile $\rho$, step size $\alpha$
   \STATE {\bfseries Output:} max reward $r_{\text{max}}$
   \STATE Initialize sampling distribution $p(x)$ to random.
   \STATE Initialize maximum reward to $r_{\text{max}}=-\infty$
   \FOR{$t=1$ {\bfseries to} $n$}
   \STATE $\mathcal{D}_{s}^{t}\leftarrow N$ samples without replacement from $p(x)$
   \STATE $\mathcal{R}_{s}^{t}\leftarrow$ Evaluate $R(x)$ for $x\in \mathcal{D}_{s}^{t}$.
   \STATE $r_{\text{max}} = max(r_{\text{max}}, max(r\in \mathcal{R}_{s}^{t}))$
   \STATE $q\leftarrow$ Calculate the $1-\rho$ quantile of $\mathcal{R}_{s}^{t}$
   \STATE $\hat{p}(x) = arg\,max_{p\in\Theta} \sum_{x\in \mathcal{D}_{s}^{t}}p(x)\mathds{1}[R(x)>q]$
   \STATE $p(x) = (1-\alpha)p(x) + \alpha\hat{p}(x)$
   \ENDFOR
   \STATE {\bfseries return} $r_{\text{max}}$
\end{algorithmic}
\end{algorithm}

Algorithm \ref{alg:cem_free} outlines the application of the Cross-Entropy Method (CEM) for optimizing black-box functions. The process begins by initializing the sampling distribution $p(x)$ to a random distribution (line 3). In each iteration, $N$ samples are drawn from the current distribution and evaluated using the ground truth function $R$ (lines 6, 7). Based on these evaluations, the distribution is then updated to concentrate more on samples that exceed the $(1-\rho)$ quantile of the $N$ rewards, i.e. focusing on the top $N \cdot \rho$ performing samples (lines 9-11). To maintain randomness and prevent overfitting, the distribution is modified by a step size $\alpha$, which typically ranges from 0.7 to 0.9. This algorithm is noted for its rapid convergence rate and robustness against different hyperparameters, as introduced in ~\cite{mannor2003cross, de2005tutorial}.

To tailor Algorithm \ref{alg:cem_free} for HLS design optimization, the key tasks involve modeling the distribution of design configurations and implementing updates to this distribution. In HLS, each design is characterized by a combination of different pragma values. Each pragma within the program must either be assigned a specific value or be set to a default value that effectively disables it, thereby completing the design configuration. We model the distribution over all possible designs by assigning a probability vector $p_i(x)$ for each pragma $i$, resulting in $p(x)=\Pi_i p_i(x_i)$. The independent modeling reduces the likelihood of overfitting when the number of data points available for distribution calculation is limited. We will demonstrate that this method of distribution modeling not only simplifies sampling distribution updates but also yields effective optimization results for HLS designs.

To implement the distribution update, we first get the one-hot embedding of each design. Specifically, for each design $x$, we use a one-hot vector $x_i$ for each pragma, where $x_{ij}=1$ if the discretized value of pragma $i$ is $j$, and $x_{ij}=0$ otherwise. For example, consider a design space containing two pragmas $P_a$ and $P_b$, and suppose $P_a$ can take value from $[1, 2, 5]$ and $P_b$ can take value from $[1,2]$. Then the design $x=\{P_a=5, P_b=2\}$ will be encoded with $\{x_a=[0,0,1], x_b=[0,1]\}$.

Then, at each iteration, the distribution $\hat{p}$ in Algorithm 1 can be calculated by separately calculating the probability vector $p_i$ of each pragma $i$, with the one-hot vector $x_i$:

\begin{equation}\label{equ:dist_update}
  \hat{p_i}=\frac{\sum_{x\in \mathcal{D}_{s}^{t}}x_i\cdot \mathds{1}[R(x)>q]}{\sum_{x\in \mathcal{D}_{s}^{t}}\mathds{1}[R(x)>q]}
\end{equation}

where $R$ is the ground truth reward function, $\mathcal{D}_{s}^{t}$ denotes the $N$ samples generated by the previous distribution $p$ at iteration $t$, and the $q$ is the $1-\rho$ quantile of $G_r$.

When implementing the algorithm, we also memorize the best $\rho\cdot N$ design in all previous iterations and consider them when calculating the quantile and optimizing the distribution, which improves the performance. This detail is omitted in the algorithm description for simplicity.

\begin{figure}
    \centering
    \includegraphics[width=0.65\columnwidth]{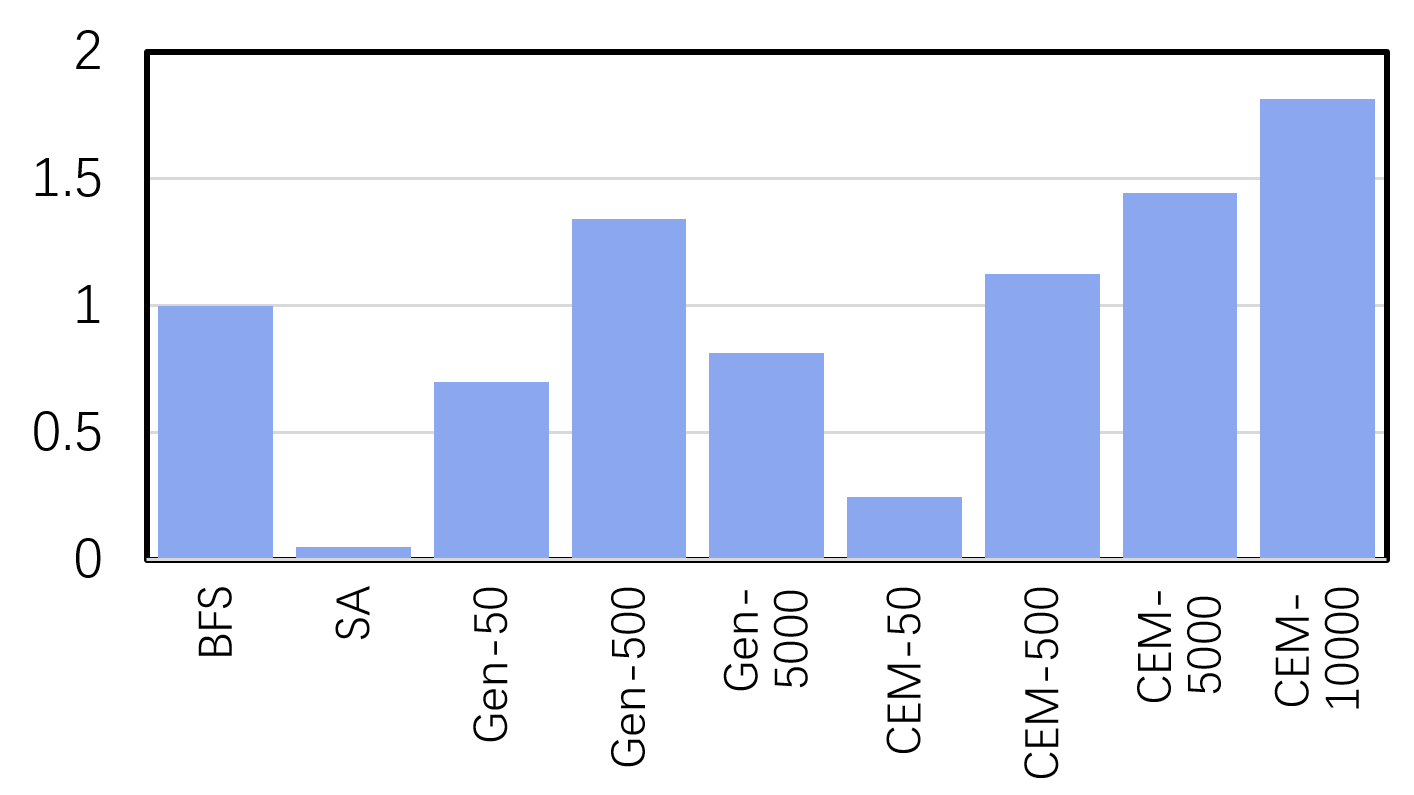}
    \caption{Normalized geomean performance of CEM and other optimization algorithms: The Gen-$K$ and the CEM-$K$ represent the different population size used in the genetic search and CEM} 
    \label{fig:cem_pred}
\end{figure}

\begin{figure}
    \centering
    \subfloat[CEM-5000]{
        \includegraphics[height=0.2\textheight]{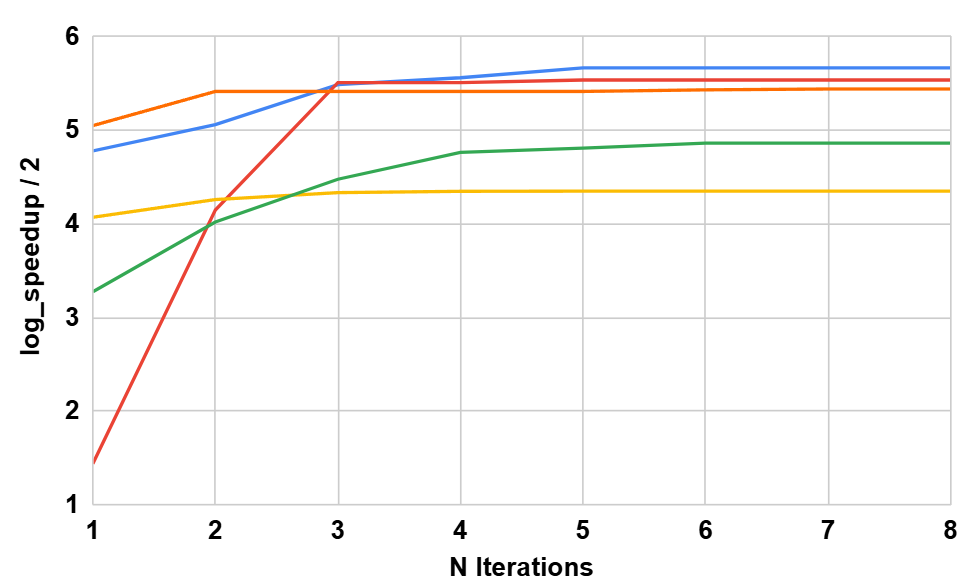}
        \label{fig:cem_converge_5}
    }
    \\
    \subfloat[CEM-10000]{
        \includegraphics[height=0.2\textheight]{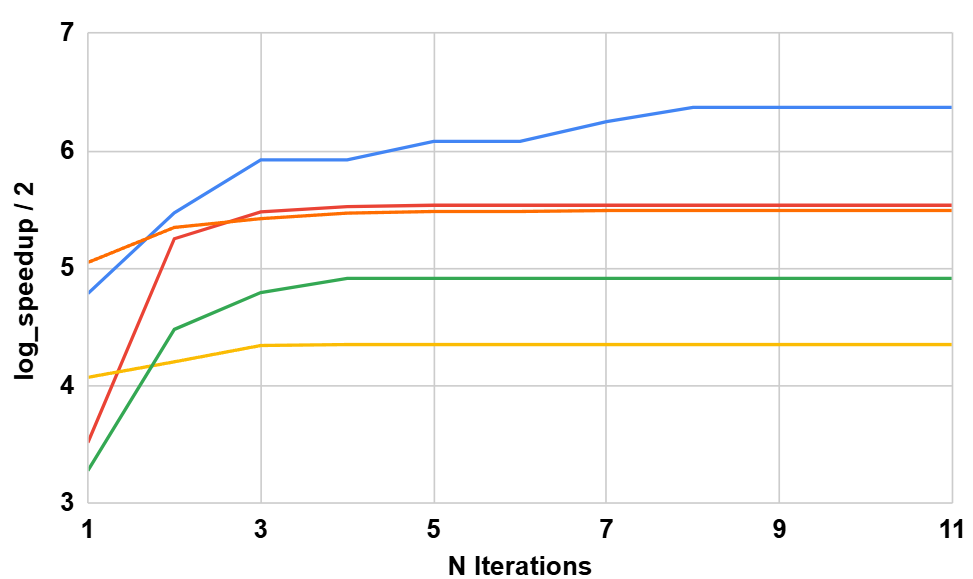}
        \label{fig:cem_converge_10}
    }
    \caption{The convergence rate of CEM-K. The horizontal axis shows the number of iterations, and the vertical axis depicts the logarithmic speedup. The different lines illustrate different kernels. The performance typically reaches a plateau in less than 10 iterations.}
    \label{fig:cem_converge}
\end{figure}

We first conduct experiments to validate that CEM is good for global optimization, comparing it with other algorithms including breadth-first search (BFS), simulated annealing, and genetic algorithm. Due to the long runtime of evaluating HLS design, we use the model prediction as the ground truth function. Even though high predicted performance may not result in high actual performance, it is a good indicator of the ability of the optimizer. We evaluate $5$ programs with a large design space and calculate the geomean of the top-1 latency under resource constraints. We also test CEM and genetic search with different population sizes. Fig. \ref{fig:cem_pred} displays the normalized result with respect to BFS. Gen-$K$ and CEM-$K$ stand for the genetic algorithm and CEM under different population sizes. Under the same runtime budget, the CEM explorer performs the best among all baselines, and the genetic search also achieves a comparable performance. When the population size is larger, the genetic search will run fewer iterations under the same time limit, resulting in the performance degradation from Gen-$500$ to Gen-$5000$. On the other hand, CEM could uncover a better design when increasing the population size, due to its fast convergence rate. As illustrated in Fig. \ref{fig:cem_converge}, the performance typically converges after $8$ iterations. However, the result also shows that optimization with CEM is not sample-efficient. It would require a very large population size, typically larger than $1000$, to reach a stable performance. When the population size $N$ is small, the distribution may quickly overfit to a specific design. Evaluating more than $1000$ designs using HLS is very time-consuming, even if a dozen of designs can be evaluated in parallel. So, we consider inserting model prediction in the loop to improve the sample efficiency of the CEM in the next section.

\subsection{Model-based CEM}
One simple way to achieve sample efficiency is to replace the evaluation on the true reward function $R$ with a proxy model $\hat{R}_\theta$, pick the design with the top prediction value, and validate it with $R$ at the last iteration. However, such an approach may cause severe performance degradation, because the model $\hat{R}_\theta$ is not accurate in terms of the label shift caused by changing the toolchain, and the distribution shift caused by a sampling distribution update. In Section \ref{sec:intro}, we illustrate how the model could be vulnerable to the label shift. Another factor to consider is the distribution shift. Even if the model has already adapted to the label shift with data on the new version, it may have low accuracy on some designs visited by the explorer, because of the mismatch between the training data distribution induced by the data gathering heuristic, and the testing distribution induced by the explorer. Therefore, it is possible for the explorer to find a design with a high predicted performance but a low actual performance. Such a phenomenon is constantly observed when trying to solely rely on the proxy model to do the optimization ~\cite{janner2019trust, gao2023scaling}. Intuitively, a powerful optimization algorithm will put pressure on the model to be perfectly accurate on the whole design space, which is usually hard to achieve given limited labeled data.

Therefore, it is critical to first update the model to have a reasonably accurate prediction under the current function $R(x)$ and the current data distribution $p(x)$. Typically, we want to optimize the following objective, where $\mathcal{D}$ is the whole design space.

\begin{align}\label{equ:min_err}
\min_{\theta} \sum_{x\in\mathcal{D}}p(x)(R(x)-\hat{R}_{\theta}(x))^2
\end{align}

When the sampling distribution updates, the objective becomes a weighted error, assigning greater weight to regions that potentially contain high-quality designs. Although it is still sample inefficient to minimize the error of the model with arbitrary data distribution on the whole design space, we find that it is possible to achieve better sample efficiency by integrating the model update into the optimization process. For the CEM, if the model can have an accurate prediction on $\mathcal{D}_{s}^t$ in each iteration, then the next distribution can be calculated solely based on the model prediction. Note that $\mathcal{D}_{s}^t\subset \mathcal{D}$ is sampled from the distribution $p(x)$, and is an unlabeled pool of the design. As discussed in section \ref{sec:cem_free}, the size of $\mathcal{D}_{s}^t$ could still be large, so we want to further improve the sample efficiency, by selecting several designs from $\mathcal{D}_{s}^t$ to update the model. The sub-problem to solve then is: How to select a few designs in $\mathcal{D}_{s}^t$ and update the model on the true value of these designs, so that the model has a minimal error on $\mathcal{D}_{s}^t$. More specifically, we want to find $\mathcal{D}_{core}^t\subset \mathcal{D}_{s}^t$, so that the model can have minimal error on $\mathcal{D}_{s}^t$ when the parameter $\theta$ is updated on $\mathcal{D}_{core}^t$.

\subsubsection{Selective Labeling for Model Update} 
Several heuristics have been proposed to tackle the data selection problem in active learning. Those heuristics primarily consider two objectives, uncertainty and diversity. Both objectives are important to minimize the model error given a fixed labeling budget.

However, the uncertainty-based approaches can only characterize the distribution shift, the change of $p(x)$. In task transfer, it is also important to be aware of the change in the function $R(x)$, or $p(y|x)$. While it is hard to estimate the change without actually observing the true value, it is still reasonable to sample diverse designs under a fixed budget, so that more information can be revealed about the unknown function. Therefore, we choose to use the coreset approach~\cite{sener2017active} to acquire diverse data.

The coreset based sampling algorithm approximates the model error on the whole pool of data with distance measurement from selected data points and the rest of the data. ~\cite{sener2017active} and ~\cite{lin2018data} separately model the problem as K-Center and K-Means clustering. We consider utilizing the K-Means algorithm on the hidden representation due to its efficiency. Since we have both a regression model $\hat{R}_{\theta_r}$ to predict the QoR and a classification model $\hat{R}_{\theta_c}$ to predict the validity of the design, we need samples to provide information for both models. We thus divide the $K$ labeling budget into $K_1$ and $K_0$ and divide the unlabeled pool $\mathcal{D}_{s}^t$ into two subsets $\mathcal{D}_{s_1}^{t}$ and $\mathcal{D}_{s_0}^{t}$. $\mathcal{D}_{s_1}^{t}$ contains designs that the classification model $\hat{R}_{\theta_c}$ predicts valid, and $\mathcal{D}_{s_0}^{t}$ contains those that the model predicts invalid. Then, we run K-Means to get $K_1$ cluster centroids of the hidden embeddings of $\hat{R}_{\theta_r}$ on $\mathcal{D}_{s_1}^{t}$ and $K_0$ cluster centroids of the hidden embeddings of $\hat{R}_{\theta_c}$ on $\mathcal{D}_{s_0}^{t}$. Then, the designs closest to the cluster centroids are selected to be added to the database.

\subsubsection{Active-CEM}
We now describe the complete exploration algorithm which consists of an outer loop that updates the sampling distribution according to the model prediction, and an inner loop that fits the model to the current distribution. We abbreviate the name with Active-CEM.

\begin{algorithm}[tb]
   \caption{Active-CEM}
   \label{alg:mbe}
\begin{algorithmic}
   \STATE {\bfseries Input:} num iterations $n$, population size $N$, true function $R$, quantile $\rho$, old model $\hat{R}'_{\theta_r}$, $\hat{R}'_{\theta_c}$, active learning round $m$, batch size $b$
   \STATE {\bfseries Output:} best design $d_{\text{best}}$, labeled database $B$
   \STATE Initialize distribution $p(x)$ to random.
   \STATE $\hat{R}_{\theta_r}, \hat{R}_{\theta_c}\leftarrow\hat{R}'_{\theta_r}, \hat{R}'_{\theta_c}$
   \FOR{$t=1$ {\bfseries to} $n$}
   \STATE $\mathcal{D}_{s}^{t}\leftarrow N$ samples without replacement from $p(x)$
   \STATE \textcolor{blue}{\# actively update model}
   \FOR{$j=1$ {\bfseries to} $m$}
       \STATE $\mathcal{D}_{core}^t\leftarrow$ select $b$ samples from $\mathcal{D}_{s}^{t}$ \hfill\textcolor{blue}{\# K-Means}
       \STATE $B=B\cup (\mathcal{D}_{core}^{t}, R(\mathcal{D}_{core}^{t}))$ \hfill\textcolor{blue}{\# run HLS on each instance in $\mathcal{D}_{core}^t$, add to $B$}
       \STATE $\hat{R}_{\theta_r}, \hat{R}_{\theta_c} \leftarrow$ train\_model($B$, $\hat{R}_{\theta_r}$, $\hat{R}_{\theta_c}$)
   \ENDFOR   
   \STATE \textcolor{blue}{\# use model prediction to update sampling distribution}
   \STATE \textcolor{blue}{\# invalid points assigned to $-\infty$}
   \STATE $\mathcal{R}_{s}^{t}(x)=\hat{R}_{\theta_r}(x)+\infty\cdot(\hat{R}_{\theta_c}(x)-1)$ for $x\in \mathcal{D}_{s}^{t}$ 
   \STATE $q\leftarrow$ Calculate the $1-\rho$ quantile of $\mathcal{R}_{s}^{t}$
   \STATE Update $p(x)$ according to $\mathcal{R}_{s}^{t}$ and $q$
   \ENDFOR
   \STATE {\bfseries return} best design found, $B$, ($\hat{R}_{\theta_r}$, $\hat{R}_{\theta_c}$)
\end{algorithmic}
\end{algorithm}

As shown in Algorithm \ref{alg:mbe}, the sampling distribution $p$ is initialized to random, and the model $\hat{R}_{\theta_r}$, $\hat{R}_{\theta_c}$ is initialized with the old model trained on the old database. A database $B$ that has designs with true labels is initialized to empty. For each iteration $t$, an unlabeled pool $\mathcal{D}_{s}^{t}$ will be sampled according to distribution $p$. Then, the unlabeled pool $\mathcal{D}_{s}^{t}$, the database $B$, and the current model will be forwarded to the inner active learning loop with the K-Means algorithm. The active learning loop will return an updated model. Then, the updated model will predict the label for all designs in $\mathcal{D}_{s}^{t}$. According to the predicted label, the $1-\rho$ quantile of the design performance will be calculated on the valid designs that satisfy the resource constraint, and the distribution will be updated according to Equation \ref{equ:dist_update}. The only difference is that the true value $R(x)$ in equation \ref{equ:dist_update} is replaced with the updated prediction model $\hat{R}_{\theta_r}(x)$ and $\hat{R}_{\theta_c}(x)$.

\subsubsection{Sample efficiency and runtime of the algorithm}
We briefly discuss the sample efficiency and the runtime of the proposed algorithm. By sample efficiency, we mean the number of designs evaluated with the HLS toolchain during exploration, which is the major runtime bottleneck. As shown in algorithm 2, the model-free version of the CEM algorithm needs to sample $N\cdot n$ design points, while the model-based version only requires $m\cdot b\cdot n$. As for the runtime, since the design is evaluated in batch, the runtime is bounded by the number of parallel HLS workers and the number of samples. The overhead of sampling unlabeled design and updating the distribution are negligible. Suppose there are $C$ parallel workers to run HLS, and the timeout for running HLS design is $T$, then the runtime of the model-free version is $Tn\cdot\left\lceil\frac{N}{C}\right\rceil$, while the runtime of the model-based version is $Tnm\cdot\left\lceil\frac{b}{C}\right\rceil$. While proper scheduling could reduce the overall runtime of evaluating $N$ design with $C$ workers, we provide an upper bound of the runtime here.

\subsection{Design Space Pruning}
\label{sec:prune}
Another way to achieve sample efficiency is to prune away part of the large design space so that the explorer will have less pressure to explore a smaller design space. While design space pruning is not the focus of this work, we empirically find that the ``TILE'' pragma is not as efficient as the ``PARALLEL'' and ``PIPELINE'' pragma for optimizing design. Also, whether or not ``TILE'' pragmas would be effective is determined mostly by the program itself (whether it is memory bound/compute bound), and is more agnostic to the change in the toolchain. Therefore, we consider pruning away ``TILE'' pragmas for some programs. We adopt a conservative heuristic that prunes away ``TILE'' pragmas only if (1) less than $10\%$ of the designs in the old database have tiling pragma, (2) the top-10 designs do not contain ``TILE'' pragma, and (3) the memory footprint of the workload could fit on-chip. We only cut the ``TILE'' pragma if all these conditions are satisfied. With this heuristic, the model will also witness less distribution shift at the early stage of exploration.

%% file: 6_evaluation.tex
\section{Evaluation}
\subsection{Experiment setup}

\begin{table}
\caption{Program domain and design space size}
\label{tab:program}
\centering
\begin{tabular}{||c | c | c||}
 \hline
 Program & Domain & DS Size \\
 \hline\hline
 2mm        & multi-MM & 4.9E+08 \\
 3mm        & multi-MM & 1.8E+13 \\
 atax       & MV & 2.6E+03\\
 atax-med   & MV & 9.5E+03\\
 covariance & Data Mining & 9.7E+08\\
 fdtd-2d    & Stencil & 1.8E+10\\
 gemm-p     & MM & 4.1E+05\\
 gemver     & MV \& VV & 1.0E+11\\
 gemver-med & MV \& VV & 1.4E+10\\
 jacobi-2d  & Stencil & 7.6E+06\\
 symm-opt   & Symmetric MM & 7.3E+05\\
 trmm-opt   & Triangular MM & 5.0E+04\\
 syr2k      & Symmetric rank-2k & 3.4E+05\\
 \hline
\end{tabular}
\end{table}

We evaluate our approach on two task transfer scenarios: from AMD Xilinx Vitis 2020.2 (V20) to Vitis 2021.1 (V21) and from Vitis 2021.1 (V21) to Vitis 2023.2 (V23). We also utilize the AMD Xilinx Merlin Compiler~\cite{cong2016source}. We assess our methods on 13 programs from the HLSyn~\cite{bai2023towards} dataset, from toolchain V20 to V21. We do not consider programs that have a very small design space. The workload covers dense linear algebra, data analytics, and stencil operations. In Appendix \ref{append:exp} we also provide the evaluation results on the complete set of HLSyn database. To evaluate the robustness of the method to different label shifts, we further test it by transferring from V21 to V23, on 5 programs with very large design space. We list the domain and the size of the design space for each program in Table \ref{tab:program}. ``MM'' stands for matrix-matrix multiplication, and ``MV'' stands for matrix-vector multiplication. Since we focus on assessing task transfer to new toolchains, we fixed the evaluation platform to Xilinx U200 with a frequency of 250MHz. 

While the proposed model architecture introduces extra overhead, we simply finetune the HARP model during the CEM exploration. After collecting a database with the explorer, we train the new model architecture with invariant embedding on the database and conduct one last round of DSE with a BFS search. The model is trained with a batch size of 64 and is trained for 1500 epochs, and the learning rate is set to 1e-3 with a linear warmup and is cosine annealed to 1e-5. When training the model with invariant embedding, we randomly sample mini-batches of 64 from the V18 and the V20 database, and forward all three batches in a single iteration to calculate the loss. We run each experiment three times with different random seeds, under the same split of the dataset, and report the mean and standard deviation of each method.

\subsection{Model accuracy}
We validate the effectiveness of learning invariant embedding on the HLS designs with model accuracy. We compare the proposed model (Inv-Hybrid) with three baselines, the HARP model trained from scratch (HARP scratch) on the new version data, the HARP model initialized with pre-trained parameters on old data and fine-tuned on new data (HARP finetune). For the old version data, we utilize the HLSyn database, which consists of data collected by AutoDSE on V18 and V20 of the toolchain. For the new version data, we utilize the data collected by AutoDSE on the V21 toolchain for the 13 programs we selected, following the data collection pipeline as described in HLSyn. 3975 regression data is gathered for the 13 programs on the V21 toolchain. And the V18 and V20 databases contain 9439 and 5194 data. We split the V21 data into training, validation, and test set, with a portion of 40\%, 10\%, and 50\%, to mimic the scenario where a few data is gathered on the large design space. We select the model with a minimum validation error and record the testing accuracy.

\begin{table}
\caption{V21 transfer learning model accuracy}
\label{tab:model_acc}
\centering
\begin{tabular}{||c c c||}
 \hline
 Model & Validation RMSE & Test RMSE \\ [0.5ex] 
 \hline\hline
 HARP Scratch & 0.361 (0.042) & 0.493 (0.031)\\
 \hline
 HARP Finetune & 0.332 (0.014) & 0.481 (0.013) \\
 \hline
 Inv-Hybrid & \textbf{0.285 (0.031)} & \textbf{0.432 (0.006)} \\
 \hline
\end{tabular}
\end{table}

Table \ref{tab:model_acc} summarizes the efficiency of each model, using the regression RMSE as a metric. We observe that the HARP model performs much better when initialized with pre-trained parameters, which is consistent with the observation made previously in~\cite{sohrabizadeh2023robust}. With the Inv-Hybrid model, the RMSE on the test set improves by $11\%$ and shows better stability during training with less standard deviation. This validates our assumption that we should model both the invariant and the difference between toolchains to achieve good transfer learning accuracy. The modeling accuracy on the full HLSyn database is reported in appendix \ref{append:acc}.

\subsection{DSE Performance}
\label{sec:eval_dse}
We next verify the efficiency of the Active-CEM algorithm. We compare with AutoDSE and HARP. For the AutoDSE baseline, we follow previous work and run all three explorers in parallel. Following~\cite{sohrabizadeh2023robust}, we set a timeout of 24 hours and do not terminate the jobs that are already running, resulting in a maximum runtime of 25 hours. For baseline evaluation, we adopt the same approach as HARP for running its explorer. Specifically, after collecting data with AutoDSE, we initialize the model with the pre-trained parameter on the old version, and then fine-tune the HARP model on the new version data. Then we run a BFS search for 1 hour with HARP and validate the design with top-10 predicted performance using the HLS toolchain. When transferring to V21, we initialize the model with parameters pre-trained on the V20 database. When transferring to V23, we initialize the model with parameters pre-trained on the V21 database. For a fair comparison, when pretraining the model on existing versions of databases, we only utilize the database collected by AutoDSE. For Active-CEM, similar to HARP, we initialize the model pre-trained parameter on the old version data. For each program, we run Active-CEM for 8 iterations to converge. At each iteration, we draw 5000 samples from the current distribution, and query 30 labeled designs among the 5000 samples. We gather all the labeled data queried so far and finetune the model for 30 epochs in each iteration. The overall runtime of the algorithm is in between $8.9-9.1$ hours. It is composed of a $8\cdot65$ minutes overall timeout for running HLS and the overhead of the policy and the model update. The runtime is $2.7x$ faster than AutoDSE. The variation in runtime is caused by generating graph data of different sizes and fine-tuning the model.

\begin{table}
\caption{V21 DSE Speedup \& Sample Efficency. We use abbreviation of the baselines. A: AutoDSE; A+H: AutoDSE+HARP; Ac-C: Active-CEM; Ac-C+H: Active-CEM+HARP; Ac-C+Inv: Active-CEM+Inv-Hybrid}
\label{tab:cem_v21}
\centering
\begin{tabular}{||c | c c c c c||}
 \hline
 Kernel & A & A+H & Ac-C & Ac-C+H & Ac-C+Inv \\
 \hline\hline
 average & 1.00 & 1.98 & 2.35 & 2.38 & 2.37\\
 geomean & 1.00 & 1.26 & 1.44 & 1.48 & 1.46\\
 \hline
 sample valid & 3975  & - &  2215 & - & -\\
 sample total & 17743 & - & 3081 & - & -\\
 \hline
\end{tabular}
\end{table}

\begin{table}
\caption{V23 DSE Speedup \& Sample Efficiency}
\label{tab:cem_v23}
\centering
\begin{tabular}{||c | c c c c||}
 \hline
  & A & A+H & A-10h & Ac-C \\
 \hline\hline
 average & 1.00 & 1.02 & 0.93 & 1.30\\
 geomean & 1.00 & 1.02 & 0.93 & 1.27\\
 \hline
 sample valid & 2786  & - & 1356 & 803 \\
 sample total & 10524 & - & 3904 & 1189\\
 \hline
\end{tabular}
\end{table}

Table \ref{tab:cem_v21} and Table \ref{tab:cem_v23} illustrate the result of the design performance when transferring from V20 to V21 and from V21 to V23. When transferring to V21, exploration with Active-CEM performs $2.34\times$ better than AutoDSE and $1.18\times$ better than the best of AutoDSE and HARP and achieves this with a $5.75\times$ total sample efficiency and a $1.74\times$ sample efficiency in terms of valid data. Note that Active-CEM tends to explore more valid designs during policy updates while AutoDSE does not contain heuristics to explicitly sample valid designs, therefore a larger portion of the designs explored by AutoDSE are invalid. As for design space pruning, among all 13 programs, only the design spaces of 3 programs (gemver, 3mm, and fdtd-2d) are pruned without searching for "tile" pragmas. We leave learning a more robust design space pruning strategy to future work. When transferring to V23, Active-CEM still performs $1.3\times$ better than AutoDSE, and $1.27\times$ better than the best of AutoDSE and HARP, and is a $8.85\times$ sample efficiency in total and a $3.47\times$ sample efficiency in terms of valid design, demonstrating the robustness of the proposed method. Better design performance and better sample efficiency are achieved by inserting model prediction into an importance sampling method, utilizing the knowledge learned on the old toolchains and addressing the label and distribution shift.

We further train the HARP model and the Invariant-hybrid model with the database collected by the CEM method, and conduct one last round of DSE with a BFS search. Our result in Table \ref{tab:cem_v21} reveals that we can locate a better design in one last round of the exploration, achieving another moderate $1.01x$ speedup on average.

We further perform an ablation study by running AutoDSE for the same amount of time as Active-CEM. Specifically, we pick the checkpoint of AutoDSE at $10$ hours and report the best design found by all three explorers. As seen in Table \ref{tab:cem_v23}, Active-CEM outperforms AutoDSE on average by $1.39\times$ when the runtime budget is the same.

\subsection{Ablation Studies}

\subsubsection{Model transfer by freezing parameters}
\label{sec:transfer_freeze}

\begin{table}
\caption{Model accuracy by freezing parameters}
\label{tab:freeze_acc}
\centering
\begin{tabular}{||c c c||}
 \hline
 Model & Validation RMSE & Test RMSE \\ 
 \hline\hline
 HARP FT No Frz & 0.332 (0.014) & \textbf{0.481 (0.013)} \\
 \hline
 HARP FT Frz(1) & 0.401 (0.07) & 0.512 (0.039) \\
 \hline
 HARP FT Frz(2) & 0.367 (0.026) & 0.501 (0.024) \\
 \hline
  HARP FT Frz(3) & 0.347 (0.028) & 0.509 (0.011) \\
 \hline
\end{tabular}
\end{table}

Table \ref{tab:freeze_acc} shows the result of transferring the model by fixing the first several layers of the GNN encoder. Similar to the previous experiment, we split the data with $40\%$, $10\%$, and $50\%$, conduct each experiment $3$ times with different random seeds, and report the mean and standard deviation. We observe that the model achieves the best accuracy when we do not freeze any parameter, which is different from the conclusion drawn from various transfer learning works with image data\cite{lin2018data}. We hypothesize that this is because of the large label shift caused by switching to a different toolchain, and the lack of proper pretraining on a large database that contains multiple versions of data.

\subsubsection{Separate modeling the invariance and the difference}

\begin{figure}
    \centering
    \includegraphics[height=0.1\textheight]{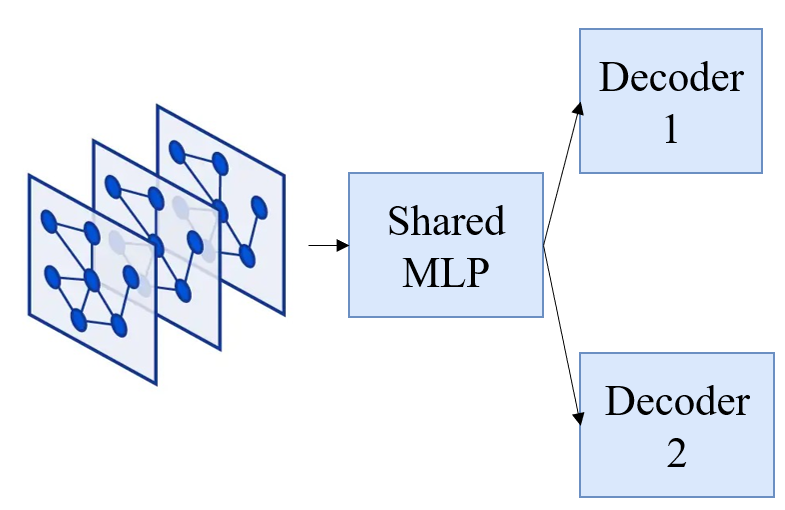}
    \caption{Inv-share: Model architecture for ablation study}
    \label{fig:model_ablation}
\end{figure}

\begin{table}
\caption{Ablation study on invariance modeling}
\vspace{-10pt}
\label{tab:abl_inv}
\centering
\begin{tabular}{||c c c||}
 \hline
 Model & Validation RMSE & Test RMSE \\ [0.5ex] 
 \hline\hline
 Inv-Share & 0.358 (0.065) & 0.481 (0.044) \\
 \hline
 Inv-Hybrid & 0.285 (0.031) & 0.432 (0.006) \\
 \hline
\end{tabular}
\end{table}

We add another baseline to confirm the effectiveness of the proposed model architecture, which models the invariance and the difference between toolchains. Specifically, we implement a model architecture as shown in Fig. \ref{fig:model_ablation}, which simply uses a different decoder to decode a shared embedding to different outputs. Moreover, we fix the parameter size of both architectures to be the same, so that we can strictly study the effect of decoupling the shared and the private embedding. In Table \ref{tab:abl_inv}, the test RMSE of ``Inv-Hybrid'' is lower than ``Inv-Share'' by $11\%$, demonstrating the importance of modeling both the invariance and the difference between toolchains.

\subsubsection{Design space pruning}

\begin{table}
\caption{Ablation study on pruning, toolchain V21}
\vspace{-10pt}
\centering
\begin{tabular}{||c c c c||}
 \hline
 Kernel & Pruned & Not Pruned & DS Size Reduction \\ 
 \hline\hline
 gemver & 1.00 & 0.37 & 15$\times$ \\
 \hline
 fdtd-2d & 1.00 & 0.92 & 39$\times$ \\
 \hline
 3mm & 1.00 & 0.88 & 244$\times$ \\
 \hline
\end{tabular}
\label{tab:abl_prune}
\end{table}

We evaluate the efficiency of design space pruning by comparing the design performance when enabling and disabling it. Only 3 out of 13 programs satisfy the pruning constraints and we report the performance difference for them. Table \ref{tab:abl_prune} illustrates how the design performance improves when design space pruning is enabled.

\subsection{Extending to Domain Transfer}

\begin{table}
\caption{Domain transfer}
\vspace{-10pt}
\centering
\subfloat[Test split 1]{
\begin{tabular}{||c c c c||}
 \hline
 Kernel & A+H & Ac-C & Ac-C+H  \\ 
 \hline\hline
 gemver-med & 1.00 & 1.73 & 1.73\\
 \hline
 fdtd-2d & 1.00 & 1.27 & 1.27\\
 \hline
 syr2k & 1.00 & 0.94 & 0.94\\
 \hline
 trmm-opt & 1.00 & 1.45 & 1.45\\
 \hline
\end{tabular}
}
\\
\subfloat[Test split 2]{
\begin{tabular}{||c c c c||}
 \hline
 Kernel & A+H & Ac-C & Ac-C+H \\ 
 \hline\hline
 3mm & 1.00 & 0.87 & 0.99 \\
 \hline
 mvt & 1.00 & 0.99 & 0.99 \\
 \hline
 covariance & 1.00 & 0.72 & 0.72 \\
 \hline
 gemm-p & 1.00 & 1.04 & 1.07 \\
 \hline
 \end{tabular}
}
\label{tab:domain}
\end{table}

We further test the efficiency of the Active-CEM algorithm by applying it in the domain transfer learning setting. Specifically, we split the HLSyn dataset, train the HARP model on the training kernels, and test it on 4 testing kernels covering stencil and linear algebra workloads. We evaluate on two different train/test split, on 8 test programs in total. While HLSyn contains variants of the same program, we ensure that the training set and the test set contain completely different programs. Table \ref{tab:domain} demonstrates that on test split 1, Active-CEM outperformed the best of AutoDSE and HARP on 3 out of 4 kernels, achieving a $2.7\times$ runtime speedup. Similarly, on test split 2, it demonstrated equal or superior performance on 3 out of 4 kernels. The performance variance across different test kernels can be attributed to the size of the design space and its difference from the training set. For instance, the training set included only one data-mining kernel, resulting in poorer performance on the covariance kernel compared to other linear algebra kernels.

%% file: 7_appendix.tex
\section{Extended experiment on the HLSyn dataset}
\label{append:exp}

\subsection{DSE Performance}
\label{append:dse}
We run Active-CEM and compare it with AutoDSE and HARP on additional kernels from the HLSyn dataset. Table \ref{tab:full_dse} presents the actual latency of the best designs found by each method and the speedup of Active-CEM over the best results from HARP and AutoDSE. Despite achieving an average speedup of $1.20\times$, Active-CEM attains greater than $2\times$ speedup on kernels such as trmm, trmm-opt, syrk, and heat-3d. It is also capable of navigating very large design spaces, achieving speedup on kernels like correlation, gemver, gemver-med, 3mm, and fdtd-2d, which have design spaces larger than $10^{10}$. However, Active-CEM's performance is worse on some kernels like mvt-med and adi. The root cause is that the model failed to learn accurate resource estimations, leading to designs that over-utilized resources or timed out on these kernels. We plan to address this issue by combining local search and global search in future work. Overall, Active-CEM can uncover better designs with less runtime on 14 out of 40 kernels.

\begin{table*}[b]
\centering
\caption{V21 DSE Latency, Speedup \& Sample Efficency. We use abbreviation of the baselines. A: AutoDSE; H: HARP; A+H: Best of AutoDSE and HARP; Ac-C: Active-CEM}
\label{tab:full_dse}
\begin{tabular}{||c | c c c c c ||}
 \hline
 Kernel & A & H & A+H & Ac-C & Speedup\\
 \hline\hline
 gemm-blocked & 2148 & 2148 & 2148 & 2148 & 1.000\\
 gemm-ncubed  & 2953 & 2657 & 2657 & 2162 & 1.229\\
 md & 4782 & 4782 & 4782 & 4782 & 1.000\\
 nw & 33906 & 34047 & 33906 & 33906 & 1.000\\
 spmv-ellpack & 7869 & 7869 & 7869 & 7869 & 1.000\\
 stencil-3d & 290528 & 290528 & 290528 & 290528 & 1.000\\
 stencil & 1372 & 1372 & 1372 & 1372 & 1.000\\
 2mm & 5389 & 4389 & 4389 & 4389 & 1.000\\
 3mm & 128908 & 9762 & 9762 & 8257 & 1.182\\
 adi & 1763481 & - & 1763481 & 7389942 & 0.239\\
 atax-med & 88117 & 92991 & 88117 & 108327 & 0.813\\
 atax & 4180 & 4180 & 4180 & 4180 & 1.000\\
 bicg-large & 173704 & 173704 & 173704 & 173704 & 1.000\\
 bicg-med & 162415 & 162416 & 162415 & 162415 & 1.000\\
 bicg & 7846 & 7846 & 7846 & 7846 & 1.000\\
 correlation & 60237 & 165135 & 60237 & 43458 & 1.386\\
 covariance & 22668 & 22168 & 22168 & 22168 & 1.000\\
 doitgen-red & 85525 & 85525 & 85525 & 85525 & 1.000\\
 doitgen & 85525 & 85525 & 85525 & 85525 & 1.000\\
 fdtd-2d-large & 2236378 & 2236378 & 2236378 & 4125477 & 0.542\\
 fdtd-2d  & 15603 & 15603 & 15603 & 11673 & 1.337\\
 gemm-p-large & 58257 & 63152 & 58257 & 67194 & 0.867\\
 gemm-p  & 9179 & 9179 & 9179 & 8827 & 1.040\\
 gemver-med & 148606 & 265686 & 148606 & 93905 & 1.583\\
 gemver  & 5685 & - & 5685 & 4875 & 1.166\\
 gesummv-med & 31985 & 31985 & 31985 & 31985 & 1.000\\
 gesummv & 4535 & 4535 & 4535 & 4535 & 1.000\\
 heat-3d & 472995 & 472995 & 472995 & 110355 & 4.286\\
 jacobi-1d  & 1668 & 1668 & 1668 & 1668 & 1.000\\
 jacobi-2d  & 164284 & 164284 & 164284 & 164284 & 1.000\\
 mvt-med  & 22644 & 22644 & 22644 & 38813 & 0.583\\
 mvt & 3982 & 3982 & 3982 & 4001 & 0.995\\
 seidel-2d & 30594317 & 30634910 & 30594317 & 30594316 & 1.000\\
 symm-opt-med & 4345927 & 35536546 & 4345927 & 5427210 & 0.801\\
 symm-opt & 13277 & 13277 & 13277 & 12646 & 1.050\\
 symm & 17478 & 17478 & 17478 & 13947 & 1.253\\
 syr2k & 45501 & 60846 & 45501 & 45501 & 1.000\\
 syrk & 45101 & 52320 & 45101 & 12999 & 3.470\\
 trmm-opt & 9387 & 7395 & 7395 & 3517 & 2.103\\
 trmm & 4091247 & 4091247 & 4091247 & 1916130 & 2.135\\
 \hline
 average & - & - & - & - & 1.201 \\
 geomean & - & - & - & - & 1.077 \\
 \hline
 sample valid & 10932 & - & - & 6626 & -\\
 sample total & 45678 & - & - & 8678 & -\\
 \hline
\end{tabular}
\end{table*}

\subsection{Model Accuracy}
\label{append:acc}

\begin{table*}
\caption{Transfer Learning Accuracy on full V21 HLSyn dataset}
\label{tab:full_acc}
\centering
\begin{tabular}{||c c c||}
 \hline
 Model & Validation RMSE & Test RMSE \\ [0.5ex] 
 \hline\hline
 HARP Scratch & 0.449 (0.071) & 0.501 (0.021)\\
 \hline
 HARP Finetune & 0.465 (0.059) & 0.498 (0.016) \\
 \hline
 Inv-Hybrid & \textbf{0.421 (0.064)} & \textbf{0.443 (0.019)} \\
 \hline
\end{tabular}
\end{table*}

We further validate the efficiency of the hybrid modeling method using the full HLSyn dataset. We use a $40\%$, $10\%$ and $50\%$ of train, validation and test split. On the full V21 dataset, there are $10932$ regression data. We compare the Inv-Hybrid approach against the HARP scratch and HARP finetune baselines. As shown in Table \ref{tab:full_acc}, the RMSE on test set improves by $11\%$.

\subsection{Overreliance on the Proxy Model}

\begin{table*}[!htp]\centering
\caption{Optimizing the HARP model with BFS and CEM: The first three columns show the top prediction performance and the prediction performance speedup of CEM over BFS, while the last three columns display the actual performance and the corresponding speedup}\label{tab:over_abl}
\begin{tabular}{|| ccccccc ||}
\hline
&BFS best prediction &CEM best prediction &prediction speedup &BFS best real &CEM best real &real speedup \\
\hline\hline
fdtd-2d-large &1143112 &80454 &14.208 &2236378 &- & \\
gemm-blocked &2114 &2114 &1.000 &2148 &2148 &1.000 \\
gemm-ncubed &2263 &2263 &1.000 &2657 &2657 &1.000 \\
md &4956 &4956 &1.000 &4782 &4782 &1.000 \\
nw &26192 &26192 &1.000 &34047 &34047 &1.000 \\
spmv-ellpack &7566 &7566 &1.000 &7869 &7869 &1.000 \\
stencil-3d &281200 &281200 &1.000 &290528 &290528 &1.000 \\
stencil &1247 &1247 &1.000 &1372 &1372 &1.000 \\
adi &1918167 &1860017 &1.031 &- &1763481 & \\
bicg-large &176127 &176127 &1.000 &173704 &173704 &1.000 \\
bicg-med &144570 &144570 &1.000 &162416 &162416 &1.000 \\
bicg &7258 &7258 &1.000 &7846 &7846 &1.000 \\
correlation &68846 &31639 &2.176 &165135 &60713 &2.720 \\
doitgen-red &82509 &82509 &1.000 &85525 &85525 &1.000 \\
doitgen &81590 &81590 &1.000 &85525 &85525 &1.000 \\
gemm-p-large &59984 &57462 &1.044 &63152 &58879 &1.073 \\
gesummv-med &31370 &31370 &1.000 &31985 &31985 &1.000 \\
gesummv &4322 &4322 &1.000 &4535 &4535 &1.000 \\
heat-3d &465158 &465158 &1.000 &472995 &472995 &1.000 \\
jacobi-1d &1567 &1567 &1.000 &1668 &1668 &1.000 \\
mvt &3676 &3466 &1.060 &3982 &3993 &0.997 \\
mvt-med &20179 &20179 &1.000 &22644 &22644 &1.000 \\
seidal-2d &27386024 &27386013 &1.000 &30634910 &30634910 &1.000 \\
symm-opt-med &536993 &536992 &1.000 &35536546 &8372367 &4.245 \\
symm &12844 &12844 &1.000 &17478 &17478 &1.000 \\
trmm &205001 &205001 &1.000 &4091247 &4091247 &1.000 \\
syrk &26025 &25358 &1.026 &52320 &15483 &3.379 \\
2mm &5521 &5458 &1.012 &4389 &5389 &0.814 \\
3mm &5257 &3415 &1.539 &9762 &- & \\
atax &4205 &4205 &1.000 &4180 &4180 &1.000 \\
atax-med &89947 &89947 &1.000 &92991 &92991 &1.000 \\
covariance &7838 &7163 &1.094 &22168 &- & \\
fdtd-2d &14651 &9238 &1.586 &15603 &32115 &0.486 \\
gemm-p &8866 &9176 &0.966 &9179 &9179 &1.000 \\
gemver &5165 &2123 &2.433 &- &- & \\
gemver-med &66471 &43038 &1.544 &265686 &3297653 &0.081 \\
jacobi-2d &109345 &109346 &1.000 &164284 &164284 &1.000 \\
symm-opt &9465 &7389 &1.281 &13277 &13277 &1.000 \\
trmm-opt &8314 &8769 &0.948 &7395 &7395 &1.000 \\
syr2k &34298 &24036 &1.427 &60846 &- & \\
\hline
\end{tabular}
\end{table*}

We validate how optimizing the surrogate model alone can lead to inferior performance. Specifically, we run the CEM algorithm with the HARP model trained on the AutoDSE database and compare it against the BFS search algorithm used by HARP, which traverses only about $100000$ points of the design space. Same to the hyperparameters setting in Section \ref{sec:eval_dse}, we run CEM with a population size of $5000$ for $8$ iterations, resulting in a sampling budget of about $40000$. After running the two search algorithms on the HARP prediction model, we select the top $10$ designs with the highest predicted performance for each kernel and run the V21 HLS tool to get the actual performance. As shown in Table \ref{tab:over_abl}, when searching with CEM, the top 1 predicted performance can be significantly higher than when searching with BFS, with notable speedup on kernels with large design spaces such as fdtd-2d-large, correlation, gemver, and 3mm. This is reasonable since CEM can search the entire design space more effectively. 

However, when validating the top 10 designs with the HLS tool, we identify a performance gap between the prediction and the ground truth. Specifically, for kernels like fdtd-2d-large, fdtd-2d, 3mm, covariance, syr2k, and gemver-medium, the top 10 designs proposed by the proxy model either contain no valid designs or have much worse actual performance than predicted. However, we also see performance improvements on some kernels like adi and syrk, suggesting variance in the model's generalizability. This set of experiments further validates that offline proxy models can be vulnerable to distribution shifts and highlights the importance of updating the model to the current data distribution as done in Active-CEM. 

Nevertheless, even when selecting the best results from CEM search, BFS search, and AutoDSE, Active-CEM still outperforms them by an average of 1.145 times. Searching with CEM on HARP only improves the result of the syrk kernel compared to the A+H baseline in Table \ref{tab:full_dse}.